\documentclass[aps,prl,
               preprint,groupedaddress,
               showpacs,
               amsmath,amssymb,amsfonts,
               graphicx,floatfix]{revtex4}

\usepackage{graphicx}                            
\usepackage{dcolumn}                             
\usepackage{bm}                                  

\newcommand{\dif}{\mathrm{d}}                    
\newcommand{\me}{\mathrm{e}}                     
\newcommand{\mpi}{\mathrm{\pi}}                  

\makeindex
\begin{document}

\title{Photon trajectories in incoherent atomic radiation trapping as
  L\'{e}vy flights}

\author{Eduardo Pereira}
\affiliation{Universidade do Minho, Escola de Ci\^{e}ncias,
  Departamento de F\'{i}sica, 4710-057 Braga, Portugal}
\email{epereira@fisica.uminho.pt}

\author{Jos\'{e} M.G. Martinho}
\author{M\'{a}rio N. Berberan-Santos}
\affiliation{Centro de Qu\'{i}mica-F\'{i}sica Molecular, Instituto
  Superior T\'{e}cnico, 1049-001 Lisboa, Portugal}

\date{\today}

\begin{abstract}
  Photon trajectories in incoherent radiation trapping for Doppler,
  Lorentz and Voigt line shapes under complete frequency redistribution are
  shown to be L\'{e}vy flights. The jump length~($r$) distributions display
  characteristic long tails. For the Lorentz line shape, the asymptotic form is a
  strict power-law~$r^{-3/2}$ while for Doppler the asymptotic
  is~$r^{-2}\left(\ln r\right) ^{-1/2}$. For the Voigt profile,
  the asymptotic form has always a Lorentz character, but the trajectory is a
  self-affine fractal with two characteristic Hausdorff scaling exponents.
\end{abstract}

\pacs{02.50.Ey, 32.80.-t, 32.70.-n}
\keywords{L\'{e}vy flight, radiation trapping, radiation
  diffusion, truncated L\'{e}vy flight, stochastic processes.}

\maketitle

%
%

Interest in distributions with long tails has increased over the
last years, with intensive search for such laws in real physical
systems. Most of the basic work on these distributions was carried
out by L\'{e}vy in the 1930s~\cite{Levy}, but it was only
relatively recently that these distributions were shown to be
applicable to the description of a number of physical, biological
and social phenomena~\cite{Bou90,Shl95,Bar02,econometrics},
including particle motion in turbulent
media~\cite{Bou90,Shl95,Shl99}, anomalous diffusion in
microheterogeneous systems~\cite{Ott1990}, chaotic transport in
laminar fluid flow~\cite{Swinney}, the albatross
flight~\cite{Nat96} and frequency fluctuations of chromophores
isolated in glassy environments~\cite{Barkai}. The purpose of this
Letter is to show that photon trajectories in incoherent radiation
trapping, a basic and common phenomenon in atomic and atmospheric
physics, and in astrophysics, fit in the category of L\'{e}vy
(superdiffusive) flights.

Although ``the general trend nowadays is to put L\'{e}vy-type
anomalous (super)diffusion on a similar footing with normal,
Brownian-type, diffusion''~\cite{Tsa1995}, the actual physical
systems invoked to justify the practical importance of L\'{e}vy
flights in physical space tend to be somewhat marginal in physical
phenomena~\cite{Shl95}. This contribution aims to prove that the
well known case of classical incoherent radiation trapping is one
of the simplest and best characterized L\'{e}vy flight found up to
now. In addition, radiation trapping has a major role in
fluorescent lamps and this could make radiation trapping the
economically most relevant concretization of a L\'{e}vy flight. In
retrospect, the~1990 Ott~\textit{et~al.}~\cite{Ott1990}
contribution on the superdiffusion in microheterogeneous systems
can no longer be considered the first experimental realization of
a random flight with infinite moments, given that it was preceded
by many studies of radiation trapping.

%
%

Radiation trapping of energy is important in areas as diverse as
stellar atmospheres~\cite{Mih}, plasmas and atomic vapors
luminescence~\cite{Mol}, terrestrial atmosphere and ocean optics,
molecular luminescence~\cite{MN}, infrared radiative transfer and
cold atoms~\cite{Bar02}. In these optically thick media, the
emitted radiation suffers several reabsorption and reemission
events before eventually escaping to the exterior; the radiation
is said to be imprisoned or trapped. Atomic \textit{radiation
trapping} is also known as \textit{imprisonment of resonance
radiation}, \textit{line transfer}, \textit{radiation diffusion}
or \textit{multiple scattering} of resonance radiation.

%
%

The first quantitative theory for atomic radiation trapping was
presented in the 1920s by Compton and Milne, that have developed a
modified diffusion equation for the (frequency) coherent spreading
of excitation. It was only in 1932 that the frequency
redistribution between absorption and reemission was taken into
account by Kenty~\cite{Ken32}. He considered a Doppler spectral
distribution and arrived to the unexpected result that, for an
infinite medium, the diffusion coefficient would be infinite. This
was the first realization of the fundamental fact that all moments
of the jump size distribution are infinite. Kenty's result shows
that a diffusion-type equation is not valid for radiation trapping
with frequency redistribution effects. Nevertheless it was only in
1947 that Hosltein and Biberman independently proposed a
Boltzmann-type integrodifferential equation~\cite{H} which remains
the starting point of the vast majority of radiation trapping
models~\cite{Mol}.

Consider the case of \textit{inelastic scattering} where, as the
result of reabsorption-reemission events there is a photon
frequency redistribution in the lab reference frame. The frequency
distribution of the emitted photons is given by the emission
spectrum~$\Theta \left( x\right)$. The absorption probability of a
photon with frequency $x$ at a given distance from the emission
point depends on the absorption spectrum~$\Phi \left( x\right)$,
and is given by~$p\left( r|x\right) =\Phi \left( x\right) \,\me
^{-\Phi \left( x\right) r}$~(Beer-Lambert law) where $r$ is the
\textit{opacity} or \textit{optical density} and is a
dimensionless distance. Radiation trapping can then be envisaged
as a random flight in physical space with spectral shape dependent
jump size distributions. The jump size distribution takes into
account the absorption probability for all possible optical
emission frequencies, hence

\begin{equation}
  \label{pr}
  p\left( r\right) =\int_{-\infty }^{+\infty }\Theta \left(
  x\right) \,p\left( r|x\right) \,\dif x
\end{equation}

Eq.~(\ref{pr}) fully characterizes the spatial aspects of the
random flight.

The moments of this distribution are

\begin{equation}
  \label{rn}
  \left\langle r^{n}\right\rangle = n! \,
  \int_{-\infty }^{+\infty } \frac{\Theta \left( x\right)}{\Phi^{n} \left( x\right)} \,
  \dif x
\end{equation}

and can be shown to be infinite for all physical reasonable atomic
emission and absorption spectral distributions as concluded by
Holstein~\cite{H}. His original analysis~\cite{H} only included
\textit{classical incoherent trapping}, in which the emitting
state is statistically unrelated with the absorbing one and there
is therefore \textit{Complete Frequency Redistribution}~(CFR). In
this case the absorption and emission spectra are identical
and~Eq.~(\ref{pr}) reduces to

\begin{equation}
  \label{prCFR}
  p\left( r\right) =\int_{-\infty }^{+\infty }
  \Phi^{2}\left( x\right) \,\me ^{-\Phi \left( x\right) r}\,\dif x
\end{equation}

Nevertheless, it can be concluded from~Eq.~(\ref{rn}) that the
case of \textit{Partial Frequency Redistribution}~(PFR) is also
characterized by an infinite moments L\'{e}vy statistics. The case
of~PFR is especially important in an astrophysical context.
Neither case of complete coherent nor complete incoherent
scattering is achieved exactly in stellar atmospheres and it is
then necessary to consider the photon redistribution and to
calculate \textit{redistribution functions} which will
give~$\Theta \left( x\right)$~\cite{Mih}. In the usual laboratory
conditions, vapor densities are high enough for~CFR to
apply~\cite{Mol}. This will be the case considered here. In
two-level~CFR atomic models both absorption and emission spectra
can be described by Doppler~\textendash ~$\Phi _{D}\left( x\right)
= \frac{1}{\sqrt{\mpi}}\me ^{-x^{2}}$~\textendash ,
Lorentz~\textendash ~$\Phi_{L}\left( x\right)
=\frac{1}{\mpi}\frac{1}{1+x^{2}}$~\textendash , or
Voigt~\textendash~$\Phi _{V}\left(x\right) =\frac{a}{\mpi
^{3/2}}\int_{-\infty }^{+\infty }\frac{\me ^{-u^{2}}}
{a^{2}+\left( x-u\right) ^{2}}\,\dif u$~\textendash ~spectral
distributions. $x$ is a normalized difference to the center of
line frequency and $a$ is the Voigt characteristic width.

%
%

We now consider the asymptotic approximations valid for large jump
sizes. A random flight in which the probability density of jump
lengths is given by

\begin{equation}\label{powerlaw}
  p\left( r\right)
  \mathrel{\mathop{\sim }\limits_{r\rightarrow \infty }}
  \frac{1}{r^{(1+\mu )}}
\end{equation}

with~$\mu<2$ is a self-similar random fractal with fractal
dimension~$\mu$. It is called a L\'{e}vy flight after
Mandelbrot~\cite{Man}, and defines a broad distribution for which
all the moments of order not smaller than $\mu $ are divergent.

If~$\Phi \left( x\right)$ is substituted for~$\Theta \left(
x\right)$ in~Eq.~(\ref{rn}) above, it is found that~$\left\langle
r\right\rangle=\infty$, whatever the spectral lineshape used, as
long as~$\Phi \left( x\right)$ is nonzero for large~$\left|
x\right|$. In this way,~$\mu \leq 1$ for any~$\Phi \left(
x\right)$.

In order to find the specific value of~$\mu $ for the spectral
distributions mentioned, we begin by rewriting~Eq.~(\ref{prCFR})
as

\begin{equation}
  \label{J1}
  p(r)=-\frac{\dif ^{2}J\left( r\right) }{\dif r^{2}}
\end{equation}

where

\begin{equation}
  \label{J2}
  J\left( r\right) =\int_{-\infty }^{+\infty }\left( 1-\me ^{-\Phi
  \left( x\right) r}\right)\,\dif x
\end{equation}

When the line shape is gaussian~(Doppler), the integrand
in~Eq.~(\ref{J2}) approaches a square wave form for large~$r$,
with inflection points at~$-x_{0}$ and~$x_{0}$ with~$\Phi \left(
x_{0}\, \right) r\simeq 1$. Hence,~$x_{0}=\sqrt{\,\ln r}$ for
large~$r$,~$J\left( r\right) \simeq 2\, x_{0} =2 \, \sqrt{\,\ln
r}$ and therefore, from~Eq.~(\ref{J1}),

\begin{equation}
  \label{powerlawDop}
  p\left( r\right)
  \mathrel{\mathop{\sim }\limits_{r\rightarrow \infty }}
  \frac{1}{r^2\left(\ln r\right) ^{1/2}}
\end{equation}

Assuming an homogeneous scaling law one arrives at an
effective~$\mu = 1+1/2\, ln(ln(r))/ln(r)$ which goes to~$1$ for~$r
\rightarrow \infty$: although the asymptotic of
Eq.~(\ref{powerlaw}) is only approximately valid, one can
nevertheless classify Doppler trapping as a strict L\'{e}vy flight
with~$\mu =1$ with all the moments of the jump distribution being
infinite.

When the line shape is Cauchy-like~(Lorentz), the integrand in
Eq.~(\ref{J2}) can be simplified to~$1-\exp \left( -\Phi \left( x
\right) \, r \right) \simeq 1-\exp \left( -\frac{r}{\mpi \, x^2}
\right)$ for large~$r$ and~$J\left( r\right)$ becomes~
$2\,\sqrt{\, r}$. Therefore,

\begin{equation}
  \label{powerlawLor}
  p\left( r \right)
  \mathrel{\mathop{\sim }\limits_{r\rightarrow \infty }}
  \frac{1}{r^{3/2}}
\end{equation}

and the asymptotic distribution is a L\'{e}vy flight with~$\mu
=1/2$.

The Voigt distribution is asymptotically coincident with the
Lorentz distribution, and therefore has also~$\mu = 1/2$ for any
value of its parameter~$a$.

Eqs.~(\ref{powerlawDop}) and~(\ref{powerlawLor}) were already
implied in Holsteins's~1947 contribution~(original equations are,
up to some minor terms, integrated versions of the equations in
this Letter), a fact hitherto unnoticed in the literature.

\begin{figure}
  \includegraphics[width=7.5cm,keepaspectratio=true]{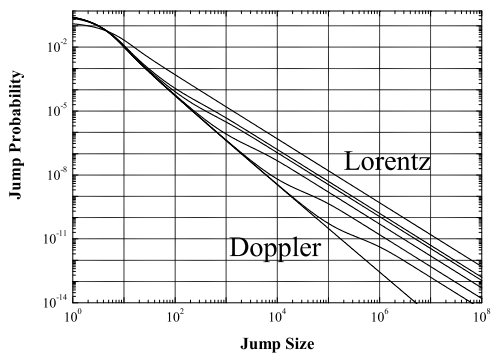}
  \caption{\label{Fig-pr}
    Jump size distribution for CFR Doppler,
    Lorentz and Voigt spectral profiles. From bottom to top:~Doppler,
    Voigt with $a=10^{-4}$, $10^{-3}$, $0.01$, $0.05$, $0.1$, and
    Lorentz.}
\end{figure}

\begin{figure}
  \includegraphics[width=7.5cm,keepaspectratio=true]{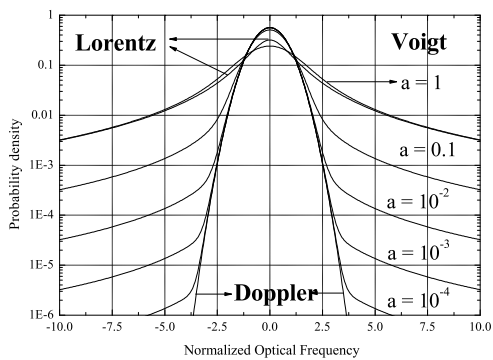}
  \caption{\label{Fig-Vgt}
    Doppler, Lorentz and Voigt spectral profiles.}
\end{figure}

Fig.~\ref{Fig-pr} show the jump probability obtained from direct
numerical integration of Eq.~(\ref{prCFR}). A linear fit gives for
the characteristic ``tail index'' (Hausdorff box counting) fractal
dimension~($\mu $) of the Doppler distribution~$\mu=1.07$~(jump
sizes $10^{3}\textendash 10^{4}$) and for the Lorentz
case~$\mu=0.500$~(jump sizes $10^{5}\textendash 10^{8}$), in
agreement with the previous assertions. Fig.~\ref{Fig-pr} also
show that the continuous transformation from Doppler into
Lorentz-like spectra as the Voigt $a$ width parameter changes from
zero into infinite values does not manifests itself in a
continuous change of the effective,~$r$ dependent,~$\mu$ values.
There is an abrupt change instead at a jump distance which scales
approximately as~$1/a$. This is expected as the asymptotic
expansion is related to the most extreme values of the wings of
the spectral distribution and these change abruptly from Doppler
to Lorentz-type asymptotics for~$a$ values as low as
$10^{-4}$~(see~Fig.~\ref{Fig-Vgt}). The data in~Figs.~\ref{Fig-pr}
and~\ref{Fig-Vgt} allows one to define the Voigt radiation
trapping trajectories as a self-affine fractal with two different
scaling exponents which manifest themselves at different
lengthscales.

%
%

\begin{figure}
  \includegraphics[width=7.5cm,keepaspectratio=true]{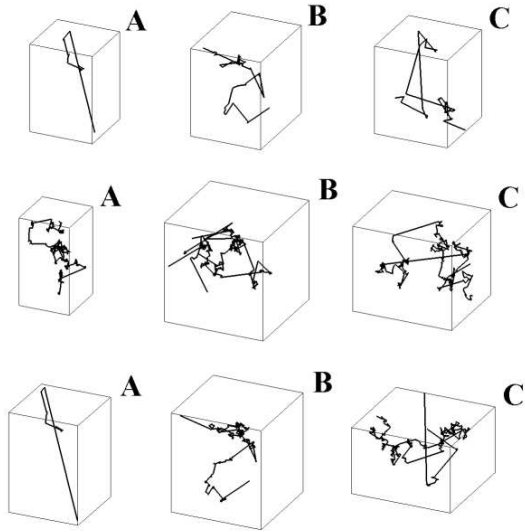}
  \caption{\label{Fig-3D}
    Single trajectories of $50\,000$ jumps each for incoherent isotropic CFR radiation
    migration with Lorentz~(top row), Doppler~(middle) and $a=0.001$ Voigt~(bottom)
    profiles in infinite~3D medium.
    Part~A show the whole trajectory while~B and~C show successive details.
    The three trajectories were obtained with the same random number sequence.}
\end{figure}

Consider now~Fig.~\ref{Fig-3D} which shows single excitation
trajectories in a 3D infinite medium~(compare Mandelbrot's well
known figure~\cite{Man}). The Lorentz and Doppler cases display
the two qualitative features characteristic of L\'{e}vy
flights:~(i)~the longer pathlength jumps, although much less
common, are of paramount importance to the overall spreading of
excitation and~(ii)~\textit{self-similar behavior}. There is an
hierarchy of clusters formed at different lengthscales but with
similar topology. The set of visited points constitute a fractal
of characteristic dimension~$\mu$ for the Doppler and Lorentz
distributions~\cite{Bou90,Shl95,Shl99}. The Voigt case shows its
\textit{self-affine} nature. The trajectory topology changes from
a Lorentz character into a Doppler one for smaller distances. The
overall topology of the trajectory as a whole is of course
dictated by the higher Lorentz scaling.

%
%

Up to now we have considered frequency redistribution at each
scattering event. However, for high opacity two-level systems
there are presumably many elastic scattering events, before an
inelastic scattering event occurs. It is therefore important to
consider at least qualitatively the influence of~\textit{elastic
scattering}. Elastic scattering events will fold up the excitation
trajectories of~Fig.~\ref{Fig-3D} and the more the higher the
ratio of elastic to inelastic scattering probabilities. From this
argument alone we will expect that an increase in the elastic to
inelastic probabilities ratio will lead to an increase in the
fractal dimension, as it will approach the~3D brownian motion with
fractal dimension~$2$.

%
%

The discussion thus far has been restricted to the case of
\textit{L\'{e}vy flights}. For the vapors studied in the
laboratory the \textit{time-of-flight} of in-transit radiation is
negligible compared to the waiting time between absorption and
reemission, thus rendering the \textit{L\'{e}vy flights} formalism
especially adequate. From the set of visited points one can obtain
spatial distribution functions of the excited density which fully
characterize the spatial aspects of radiation migration. The
temporal evolution can be factorized from the spatial
part~(L\'{e}vy flight) and then separately
handled~\cite{MN,epereira}. However, for interstellar gases the
time between scattering events might be large compared to the
absorption-reemission times. Because the speed of light is finite,
a~\textit{L\'{e}vy walk} modification of the flights here
presented will be more appropriate in an astrophysical
context~\cite{Shl99}.

%
%

The theoretical models for incoherent atomic radiation trapping
are based either on the Holstein-Biberman multi-exponential mode
expansion or on the so-called \textit{multiple scattering
representation}~\cite{Mol}. In the last case, the temporal
evolution is known analytically and the spatial distributions can
be separately obtained, either directly from the~Eq.~(\ref{pr}) or
from its asymptotic
approximation~Eq.~(\ref{powerlaw})~\cite{epereira}. It is within
this theoretical framework that the results reported in this
Letter are most useful. On the other hand, the individual terms in
the Holstein expansion have no direct relation with the one (or
$n^{th}$) step jump probabilities. Although both approaches are
ultimately equivalent~\cite{MN,Lai}, we follow the multiple
scattering approach, preferable for systems of low
opacity~\cite{MN}. Also, the multiple scattering representation
has a simple interpretation, since each term corresponds to a
specific generation of excited atoms or molecules. In a large
number of practical situations the opacity is not high enough to
warrant the exclusive use of Holstein's slowest exponential mode.
On the other hand, the multi-exponential expansion remains valid,
but its fundamental mode cannot be identified with Holstein's high
opacity result and should be estimated by the \textit{stationary
mode} associated with a non-changing spatial distribution
function~\cite{MN,Lai}, a point often misunderstood in the
literature.

%
%

Results of~Fig.~\ref{Fig-3D} strictly apply to an infinite~3D
medium but in a vast number of experimental situations the system
is finite and the trajectory is \textit{truncated} before the
asymptotic L\'{e}vy expansion is able to manifest itself. Mantegna
and Stanley~\cite{Man94} introduced in~1994 a class of L\'{e}vy
flights, the \textit{truncated L\'{e}vy flight}~(TLF), in which
the largest steps of an ordinary L\'{e}vy flight are eliminated by
a sharp cutoff in its power tail. This work allows a reassessment
of Kenty's pioneering contribution since he was the first to have
used a truncation procedure: within the framework of the kinetic
theory of gases he considered a truncated Maxwell distribution of
speeds, with a maximum speed corresponding to a free path equal to
the linear size of sample cell~\cite{Ken32}. Although
\textit{truncation of the jump size} renders the moments of the
distribution finite, the convergence to a Gaussian can be
extremely slow and the random walk can exhibit anomalous behavior
and multi-scaling properties in a wide range before
convergence~\cite{Man94,NakKop}. Moreover, \textit{trajectory
truncation} is more complex to handle and does not create
convergence to the Gaussian in usual situations.

%
%

In this Letter, it was shown that all photon trajectories arising
from incoherent two-level complete frequency redistribution
trapping are superdiffusive L\'{e}vy flights with~$\mu \leq 1$. In
particular, for the Doppler lineshape~$\mu = 1$, whereas for
Lorentz and Voigt~$\mu = 1/2$.

\begin{acknowledgments}
E.~Pereira would like to acknowledge fruitful discussions with
Fran\c{c}ois Bardou~(CNRS,~IPCMS).

This work was supported by Funda\c{c}\~{a}o para a Ci\^{e}ncia e
Tecnologia~(FCT,~Portugal) within project~POCTI/34836/FIS/2000.
\end{acknowledgments}

\end{document}